\documentclass[conference]{IEEEtran}
\usepackage{graphicx}
\usepackage{amssymb}
\usepackage{amsmath}
\usepackage{cite}
\usepackage{amsthm}
\usepackage{epstopdf}
\usepackage{color}

\usepackage{multirow}
\usepackage{booktabs}
\usepackage{amsfonts}
\usepackage{epsfig}
\usepackage{url}
\usepackage{algorithm}
\usepackage{algorithmic}
\usepackage[tight]{subfigure}

\makeatletter
\def\ScaleIfNeeded{%
\ifdim\Gin@nat@width>\linewidth \linewidth \else \Gin@nat@width \fi
} \makeatother

\DeclareMathOperator{\E}{\mathbb{E}}

\newcommand{\EXs}[2]{\E_{{#1}}\left\{{#2}\right\}}

\newcommand{\ds}{\displaystyle}

\newcommand{\clS}{{\cal S}}
\newcommand{\clA}{{\cal A}}
\newcommand{\clC}{{\cal C}}
\newcommand{\clP}{{\cal P}}
\newcommand{\clR}{{\cal R}}
\newcommand{\clM}{{\cal M}}
\newcommand{\clN}{{\cal N}}
\newcommand{\clL}{{\cal L}}

\newcommand{\clI}{{\cal I}}

\newcommand{\wei}{{\boldsymbol{\theta}}}
\newcommand{\weitar}{{\wei"}}
\newcommand{\dfactor}{\gamma}
\newcommand{\lrate}{\beta}
\newcommand{\emax}{e^{max}}

\begin{document}

\title{Resource Allocation in Mobility-Aware Federated Learning Networks: A Deep Reinforcement Learning Approach}

\author{\IEEEauthorblockN{Huy T. Nguyen\IEEEauthorrefmark{1}, Nguyen Cong Luong\IEEEauthorrefmark{1}, Jun Zhao\IEEEauthorrefmark{1}, Chau Yuen\IEEEauthorrefmark{2}, and Dusit Niyato\IEEEauthorrefmark{1}}
	\IEEEauthorblockA{\IEEEauthorrefmark{1} School of Computer Science and Engineering,
		Nanyang Technological University,
		Singapore\\
		\IEEEauthorrefmark{2} Engineering Product Development, Singapore University of Technology and Design, Singapore\\
		Email: \IEEEauthorrefmark{1} \{huyt.nguyen, clnguyen, junzhao, dniyato\}@ntu.edu.sg, \IEEEauthorrefmark{2} yuenchau@sutd.edu.sg.}}

\markboth{Prepared for Submission in IEEE Trans. on XXX}
 {Nguyen \textit{\MakeLowercase{et al.}}: XXX}

\maketitle

\thispagestyle{empty}

\begin{abstract}
Federated learning allows mobile devices, i.e., workers, to use their local data to  collaboratively train a global model required by the model owner. Federated learning thus addresses the privacy issues of traditional machine learning.  However, federated learning faces the energy constraints of the workers and the high network resource cost due to the fact that a number of global model transmissions may be required to achieve the target accuracy. To address the energy constraint, a power beacon can be used that recharges energy to the workers. However, the model owner may need to pay an energy cost to the power beacon for the energy recharge. To address the high network resource cost, the model owner can use a WiFi channel, called \textit{default channel}, for the global model transmissions. However, communication interruptions may occur due to the instability of the default channel quality. For this, \textit{special channels} such as LTE channels can be used, but this incurs channel cost. As such, the problem of the model owner is to decide amounts of energy recharged to the workers and to choose channels used to transmit its global model to the workers to maximize the number of global model transmissions while minimizing the energy and channel costs. This is challenging for the model owner under the uncertainty of the channel, energy and mobility states of the workers. In this paper, we thus propose to employ the Deep Q-Network (DQN) that enables the model owner to find the optimal decisions on the energy and the channels without any a priori network knowledge. Simulation results show that the proposed DQN always achieves better performance compared to the conventional algorithms. 
\end{abstract}

\begin{IEEEkeywords}
Federated learning, deep reinforcement learning, channel selection, energy allocation, mobility.
\end{IEEEkeywords}

\section{Introduction}\label{section:2}

In traditional machine learning approaches, neural network models are trained at a server or a data center. Thus, the centralized learning approaches typically require the raw data, e.g., photos and location information, collected by mobile devices to be centralized at the server \cite{jing2017crowdtracker}. The centralized learning approaches thus face big issues including privacy, long propagation delay, and backbone network burden \cite{mao2017survey}.  

Recently, federated learning as a decentralized machine learning approach has been proposed  to address the above issues \cite{mcmahan2016communication, anh2019efficient}. In the federated learning, mobile devices, i.e., workers, are required to collaboratively train the neural network model of the model owner\footnote{In the rest of the paper, we use ``model owner" to refer to the server which creates the global model, and ``worker" is a mobile device which trains the global model.}. In particular, the model owner first transmits its global model to the workers. The workers then use their data to train the model locally and send the model updates to the model owner. The model owner aggregates the model updates from the workers to a new global model and transmits it back to the workers for training. The model owner and the workers periodically exchange and update the model until a target accuracy is achieved \cite{mcmahan2017federated}. By updating the models rather than the raw data, the federated learning alleviates many challenging problems, e.g, privacy issues and the backbone network burden issues, of the traditional machine learning \cite{lim2019federated}. However, the federated learning faces two limitations. 

The first limitation is that the workers as mobile devices have energy constraints, and this can make the workers inactive. To address this limitation, a power beacon can be used to recharge energy to the workers. However, the model owner may need to pay an energy cost to the power beacon for the energy recharge. Thus, the model owner must decide appropriate amounts of energy recharged to the workers. The second limitation is that to achieve the target accuracy of the global model, a number of global model transmissions from the model owner to the workers may be required. This incurs a high network resource cost, i.e., the bandwidth cost, to the model owner. To reduce the network resource cost, the model owner uses the WiFi channel or Bluetooth channel, called \textit{default channel}, for the global model transmissions. As such, the model owner can transmit its global model to the workers with a free access cost. However, the default channel sometimes has low quality, and the global model transmissions may be unreliable \cite{xu2016enterprise}. Moreover, the coverage of the default channel is very short that may result in the communication interruptions between the model owner and the workers due to the mobility of the workers. The model owner can thus use so-called \textit{special channels} such as LTE channels or TV White Space channels for the global model transmissions \cite{afzal2018unlocking}. However, using such a special channel requires a high communication cost, i.e., channel cost. 

As such, the problem of the model owner is (i) to decide amounts of energy recharged to the workers and (ii) to choose channels, i.e., the default channel or the special channels, for the global model transmissions to maximize the number of successful transmissions while minimizing the energy cost and the channel cost. This is challenging for the model owner since the mobile environment is stochastic in which the channel, energy, and mobility states are uncertain and unpredictable. In this paper, we thus propose to employ the Deep Q-Network (DQN) \cite{mnih2015human} that enables the model owner to find the optimal decisions on the energy and the channels with no existing network knowledge. In particular, we first formulate a stochastic optimization problem of the model owner that maximizes the number of global model transmissions while minimizing the energy cost and the channel cost. Then, the Deep Q-Learning (DQL) algorithm with Double Deep Q-Network (DDQN) \cite{van2016deep} is adopted to derive the optimal policy for the model owner. Simulation results show that the proposed DQL always achieves the better performance compared with baseline algorithms.

\section{System Model}
\label{section:3}

\begin{figure}[t!]
	\centerline{\includegraphics[width=0.5\textwidth]{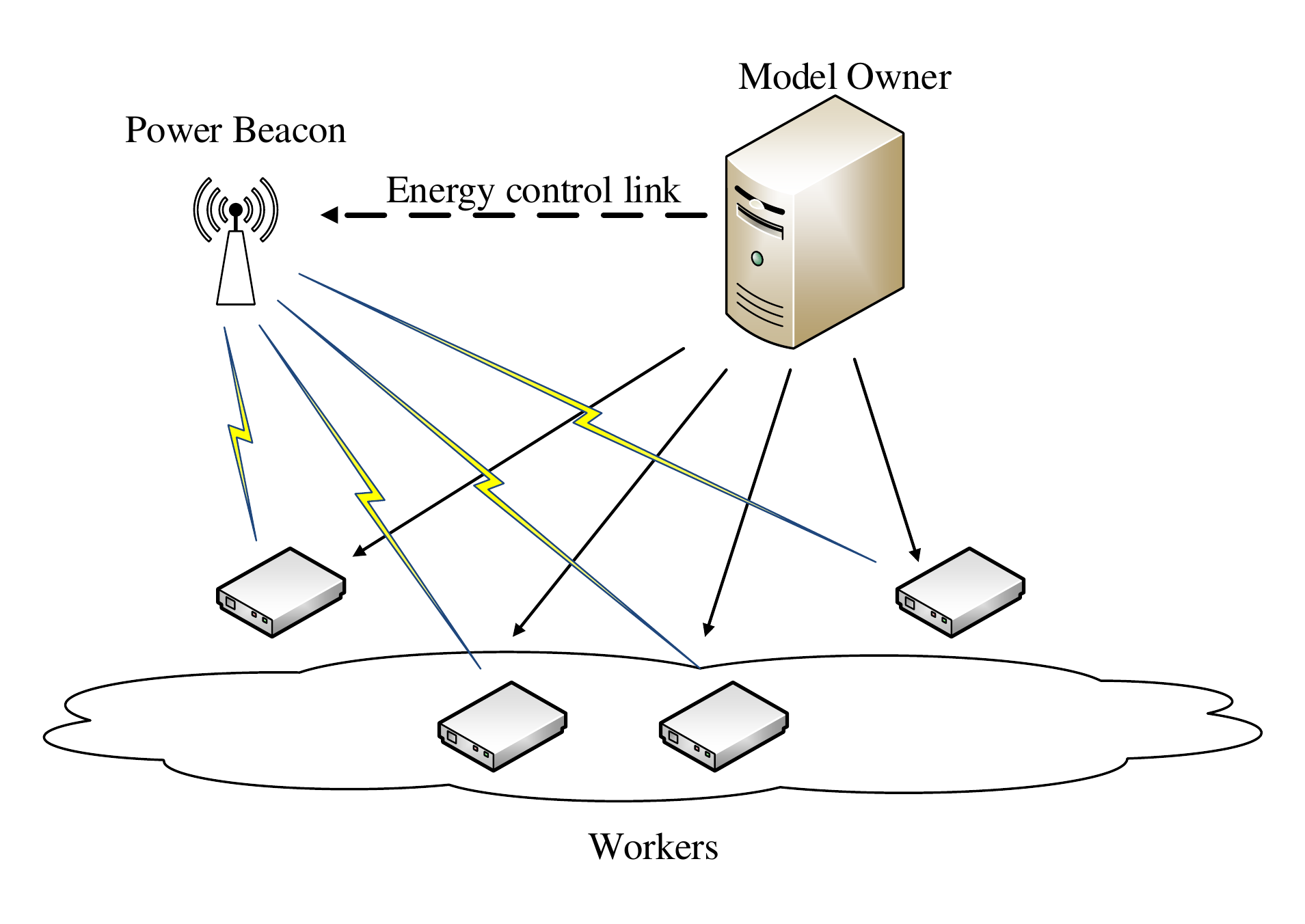}}
	\caption{A downlink model of FLNet.}
	\label{fig:model}
\end{figure}

We consider a downlink model of a Federated Learning Network (FLNet) as shown in Fig.~\ref{fig:model}. The network consists of a model owner and a set $\clL$ of $L$ workers, i.e., $\clL \triangleq \{1,\dots, L\}$. The model owner generates a global model, i.e., weight parameters, and transmits the model to the workers for training. The training requires multiple iterations until the global accuracy is achieved \cite{tran2019federated}. Each worker is equipped with a capacity-limited battery that can store a maximum number of $\emax$ energy units. Note that as there is few energy in the battery, the worker may become inactive and cannot communicate with the server. The worker's battery can be recharged by a power beacon, and the model owner pays an energy cost to the power beacon for the energy recharging. To minimize the energy cost while reducing the inactivity probability of the workers, the model owner needs to determine appropriate amounts of energy recharged to the workers, e.g., through energy control links. Note that the workers may have different distances to the power beacon and consume different amounts of energy for uploading their local models. Let $\mu_l$ denote the weighted metric of recharging energy to worker $l$. Without loss of generality, assuming that the workers with higher index have longer distances to the power beacon and higher weighted metrics, i.e., $\mu_1 \leq \dots \leq \mu_L$. As such, using $\mu_l$ is to guarantee ``fairness" among the workers. 


\begin{figure*}[t!]
	\centerline{\includegraphics[width=1\textwidth]{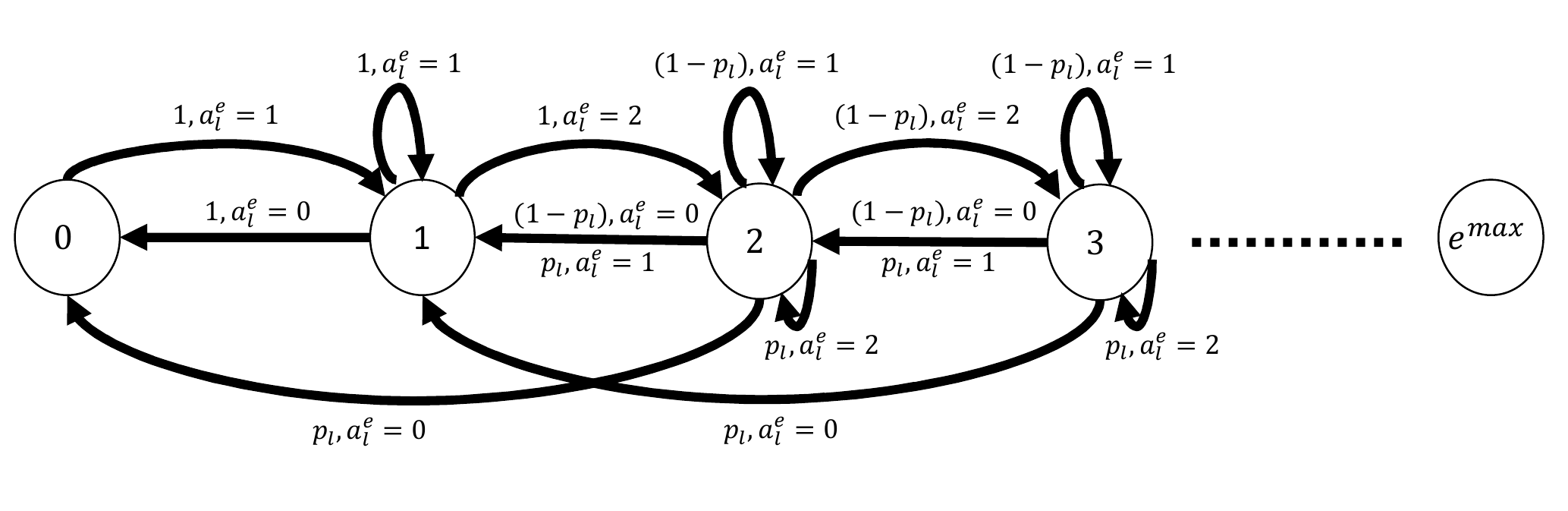}}
	\caption{Energy state transition model of worker $l$.}
	\label{fig:Emarkov}
\end{figure*}

As mobile devices, the workers have a possibility of movement that can result in the link disconnection and communication interruption as the worker is out of the coverage of the power beacon. Denote $q_l$ as the probability that worker $l$ is present in the coverage. Within the coverage, the model owner transmits its global model to the workers by using the WiFi channel or Bluetooth channel, i.e., default channel, with free-cost access. The default channel has an unstable connection due to the low quality. Let $p_{su}$ denote the probability that the model owner transmits its global model over the default channel successfully. This means that the transmission fails with a probability of $(1-p_{su})$. To mitigate the communication interruption between the model owner and the workers, special channels, e.g., the LTE channels and TV White Space channels, can be employed. Denote $\clN$ as a set of $N-1$ special channels, i.e., $\clN \triangleq \{2,\dots, N\}$. As the model owner decides to use special channel $n\in \clN$, it is charged with an access cost $\lambda_n >0$ that is proportional to the channel quality. The special channels may have different quality, and we assume that $\lambda_2 < \dots < \lambda_N$. Denote $p_{su}^n$ as the probability that the global model transmission over special channel $n$ is successful. 

The problem of the model owner is to decide which channel, i.e., the default channel or one of $N-1$ special channels, can be used to transmit its global model to each worker to maximize the number of successful transmissions and minimize the channel cost. In addition, the model owner needs to decide appropriate amounts of energy recharged to the workers to minimize the energy cost while mitigating the inactivity possibility of the workers.

\section{Problem Formulation}
\label{section:4}
Under the uncertainty of the states of the channels and the workers, the problem of the model owner can be formulated as a stochastic optimization problem that defined by a tuple $<\clS, \clA, \clP, \clR>$, where
\begin{itemize}
\item $\clS$: The state space of the network.
\item $\clA$: The action space of the model owner.
\item $\clP$: The state transition probability function, where the current state $s \in \clS$ transits to the next state $s' \in \clS$ with probability $P_{s,s'}(a)$ when action $a \in \clA$ is executed.
\item $\clR$: The reward function of the model owner.
\end{itemize}

First, we consider the action space of the model owner. Given the problem mentioned in Section II, the action space of the model owner can be expressed as follows:
\begin{align}
\mathcal{A}= &\big{\{} \left (a_1^c, a_1^e, \dots, a_l^c, a_l^e, \dots, a_L^c, a_L^e \right ); \nonumber \\
&{\color{black}a_l^c \in \{0,1,2,\dots,N\}}, a_l^e \in \{0,1,\dots, e^{max}\}, l \in \clL \big{\}}, \notag
\end{align}
where $a_l^c = 0$ means that the model owner does not transmit the global model to worker $l$, $a_l^c=1$ means that the default channel is used for transmitting the global model to worker $l$, $a_l^c=n$ means that special channel $n, n\in \clN$ is used for transmitting the model to worker $l$, and $a_l^e$ refers to the amount of energy recharged to worker $l$. 

Next, we consider the state space of the network. The state space of the network can be regarded as a combination of state spaces of $L$ workers, i.e., $\mathcal{S} = \prod_{l=1}^{L}\mathcal{S}_l$, where $\prod$ is the Cartesian product, and $\mathcal{S}_l$ is the state space of worker $l$ that is defined as follows: 
\begin{align}
\mathcal{S}_l = &\big{\{} (w_l, e_l, X_l); \nonumber \\ 
&w_l \in  \{0,1\},  e_l \in \{ 0,1,\ldots,e^{max} \}, X_l \in  \{0,1\} \big{\}}, \notag
\end{align}
where {\color{black}$w_l$ is the state of channel that the model owner uses to transmit its global model to worker $l$, $w_l=1$ if the channel is good, and $w_l=0$ otherwise.} $e_l$ refers to the energy state of worker $l$ that is the current number of energy units in the battery. $ X_l$ is the mobility state of worker $l$, $ X_l = 1 $ means that worker $l$ is in the communication range of the model owner, and $X_l = 0$ otherwise.

Now, we consider the state transition of the model owner. For the energy transition, each worker consumes its energy for training the model required by the model owner and running its local applications. In general, the amount of consumed energy is unknown to the model owner. To model the transition of energy state for the worker, we use the Markov chain. We assume that in each training iteration, the worker consumes at least one energy unit and at most two energy units for training the model and running local application. The recharging energy to the workers happens after the training in each iteration. The energy state transition of worker $l$ is shown in Fig. \ref{fig:Emarkov}, where the energy is reduced by either one unit with a probability of $(1-p_l)$ or two units with a probability of $p_l$ under the condition of $e_l\geq2$. Note that at $e_l=1$, the energy state can directly transfer to state $e_l=0$ with the probability of $1$ due to the energy consumption of running local applications. For the channel and mobility, we model the channel state $w_l$ and the mobility state $X_l$ as a Bernoulli process, which takes value $1$ with probability $p_{su}$ or $p_{su}^n, \forall n \in \clN$ and $q_l$, respectively. 

Finally, we define the reward of the model owner. One of the objectives is to maximize the number of successful transmissions of the model owner. We assume that the model owner earns a positive utility $\Delta$ from the successful transmission to each worker. Let $\clI_l$ denote the utility that the model owner receives for transmitting its global model to worker $l$. $\clI_l$ can be expressed as follows:
\[
\clI_l =\begin{cases}\begin{array}{ll}
\Delta & \mbox{if}\ \mbox{the transmission is successful}, \cr
0&\mbox{otherwise}.
\end{array}
\end{cases}
\]

As the model owner decides to use special channel $\lambda_n$ to transmit the global model to worker $l$, the model owner must pay channel access cost $\clC_l^c=\lambda_n$, i.e., to a network provider. Otherwise, as the model owner chooses not to transmit or chooses the default channel to transmit the global model to worker $l$, $\clC_l^c=0$. Thus, $\clC_l^c$ is determined as follows:
\[
\clC_l^c =\begin{cases}\begin{array}{ll}
\lambda_n & \mbox{if}\ a_l^c = n, n \in \clN, \cr
0 & \mbox{if}\ a_l^c = 0 \; \mbox{or} \; a_l^c = 1.
\end{array}
\end{cases}
\]

When the model owner decides to transmit its global model to worker $l$, i.e., by using either the default channel or the special channel, the worker consumes energy for training and uploading the local model. The worker's battery can be recharged by the power beacon, and there is energy cost $\clC_l^e$. $\clC_l^e$ is the cost that the model owner pays the power beacon for recharging energy to worker $l$. $\clC_l^e$ is a function of the weighted metric $\mu_l$ and the amount of recharging energy $a_l^e$. $\clC_l^e$ is given by
\[
\clC_l^e =\begin{cases}\begin{array}{ll}
\mu_l a_l^e & \mbox{if $X_l = 1$}, \cr
\mu_{out} a_l^e & \mbox{otherwise}, 
\end{array}
\end{cases}
\]
where $\mu_{out}$ is the weighted metric for recharging energy from outside the coverage. Note that the model owner pays a higher cost to recharge the energy if the worker is outside of the coverage, meaning that $\mu_{out} > \mu_l, \forall l \in \clL$.    

The reward of the model owner is defined as a function of state $s \in \mathcal{S}$ and action $a \in \mathcal{A}$ as follows:
\begin{equation}
\mathcal{R}\left ( s,a \right )= \ds\sum_{l\in\clL} \left( \alpha_\clI \frac{\clI_l}{\clI^{max}} - \alpha_c \frac{\clC_l^c}{\clC^{c,max}} - \alpha_e \frac{ \clC_l^e}{\clC^{e,max}} \right),
\end{equation}
where $\alpha_\clI, \alpha_c$, and $\alpha_e$ are the scale factors. $\clI^{max} = \Delta L, \clC^{c,max} = N L$ and $\clC^{e,max} = \mu_{out} e^{max} L$ are the maximum values of the total utility, channel cost and energy cost, respectively. 

Given each state $s \in \clS$, the model owner must determine the optimal action $a \in \clA$ to maximize the accumulated reward. The output is the optimal policy, which is defined as $\pi^* : {\mathcal{S}} \rightarrow {\mathcal{A}}$. To obtain the optimal policy $\pi^*$, the conventional Q-Learning (QL) algorithm \cite{watkins1992q} can be utilized. The main idea of QL algorithm is to update $Q$-values, i.e., $Q(s,a)$, of state-action pairs for a $Q$-table by using Bellman's equation as follows \cite{watkins1992q}: 
\begin{align}
Q(s,a) = \ds\sum_{s'\in\clS}  P_{\pi}(s,s') (\clR(s,a) + \dfactor V(s')).
\end{align}
Thus, the $Q$-value is updated as follows
\begin{align}
Q^{new}(s,a) = (1-\lrate) Q(s,a) + \lrate \big{(}\clR(s,a) + \dfactor \ds\max_{a'\in \clA} Q(s',a')\big{)}, 
\end{align}
where $\lrate$ is the learning rate, and $\dfactor$ is the discount factor, $0 \leq \dfactor < 1$.

After updating the $Q(s,a)$ values, the model owner can rely on the $Q$-table to determine the optimal action from any state to maximize the accumulated reward. However, this QL algorithm is only feasible for networks with small state and action spaces. As the number of workers $L$ increases, the problem of the model owner is high dimensional due to the involvement of the large state and action spaces. Therefore, the Deep Q-Learning (DQL) algorithm \cite{mnih2015human}, which is a combination of deep neural network (DNN) and QL, is adopted to find the optimal policy for the model owner.   

\section{Deep Q-Learning Algorithm}

\begin{algorithm}[!t]
	\centering
	\caption{DQL algorithm}\label{Alg1}
	\begin{algorithmic}
		\STATE \textbf{Initialization} $\dfactor$, $\lrate$, $\wei, \weitar$, $\kappa=0$
		\REPEAT 
		\STATE Initialize network state $s$ from $L$ workers, $i=0$
		\REPEAT
		\STATE Choose action $a$ according to $\epsilon$-greedy policy
		\STATE Execute action $a$ and obtain reward $r$ and next state $s'$
		\STATE Store tuple $<s,a,r,s'>$ in $\clM$
		\STATE Sample $\clM_b$ experiences $<s,a,r,s'>$ from $\clM$
		\STATE Determine $a^{max} = \arg\max_{a'\in \clA} Q(s',a';\wei)$ 
		\STATE Execute \eqref{ddqn} to obtain $y_{DDQN}$
		\STATE Perform gradient descent step on \eqref{loss} to update $\wei$
		\STATE Reset $\weitar=\wei$ every $\clI^{-}$ iterations
		\STATE Reset $i \leftarrow i + 1$
		\UNTIL $i$ is greater than maximum number of iterations in episode $\kappa$
		\STATE Reset $\kappa \leftarrow \kappa + 1$
		\UNTIL $\kappa$ is greater than maximum number of episodes
	\end{algorithmic}
\end{algorithm}

Different from the QL algorithm, the DQL algorithm uses a DNN to derive approximate $Q$-values, i.e., $Q^*(s,a)$, instead of the $Q$-table. The input of the DNN is one of states of the model owner, and the output includes $Q$-values $Q(s,a;\wei)$ of all possible actions, where $\wei$ is the weights of the DNN. To obtain the approximate values $Q^*(s,a)$, the DNN needs to be trained by using experiences $<s,a,r,s'>$. In particular, the DQL algorithm updates weights $\wei$ of the DQN to minimize the loss function defined as follows: 
\begin{align} \label{loss}
L(\wei) = \EXs{}{(y_{DQN}(t) - Q(s,a;\wei))^2},
\end{align}
where $y_{DQN}$ is the target value that is given by
\begin{align} \label{dqn}
y_{DQN} = r + \dfactor \arg \max_{a'\in \clA} Q(s',a';\wei'),
\end{align}
where $\wei'$ is the weights of the DNN from the previous iteration and $r$ is the current reward. Note that action $a$ is selected according to the $\epsilon$-greedy policy. From \eqref{dqn}, we observe that the max operator uses the same $Q$-value for both action selection and action evaluation. As a result, the derived policy may be inaccurate due to the over-optimistic estimation \cite{anh2018deep}.

To address the over-optimistic problem, the action selection should be decoupled from the action evaluation. For this, the Double Deep Q-network (DDQN) \cite{van2016deep} can be used. The main feature of DDQN is the use of two separate DNNs i.e., an online network with weights $\wei$ and a target network with weights $\weitar$. The weights of the online network are updated at each iteration, while those of the target network are kept constant. For every $\clI^{-}$ iterations, the target network's weights $\weitar$ are reset to $\wei$. The target function of DDQN is defined by
\begin{align} \label{ddqn}
y_{DDQN} = r + \dfactor Q(s',\arg \max_{a'\in \clA} Q(s',a';\wei); \weitar).
\end{align}

As seen from \eqref{ddqn}, the weights of the online network, i.e., $\wei$, are used to select an action, while those of the target network, i.e., $\weitar$, are used to evaluate the action. Both the online network and the target network use the next state $s'$ to compute the optimal value $Q(s',a';\wei)$. Given $\dfactor$ and $r$, the target value $y_{DDQN}$ is calculated based on \eqref{ddqn}. Then, a gradient descent step is performed to update the weights of online networks $\wei$ based on the loss function $L(\wei)$ in \eqref{loss}. To guarantee the stability of the learning, the DQL algorithm employs an experience replay memory $\clM$, where a mini-batch of $\clM_b$ experiences is taken at each iteration to train the DNNs. Algorithm \ref{Alg1} shows how to implement the DQL algorithm.

\section{Numerical Results}
In this section, we present experimental results to evaluate the
performance of the proposed DQL algorithm. For comparison, we use the QL~\cite{watkins1992q}, greedy, and random algorithms as baseline schemes. For the greedy algorithm, the model owner decides the maximum amount of energy charged to each worker and selects the special channel with the highest quality to transmit the global model to the worker. For the random algorithm, selecting the channel and deciding the amount of charging energy for each worker are random. The algorithms are implemented by using TensorFlow deep learning library \cite{abadi2016tensorflow}. In particular for the DQL, we employ two DNNs, and each DNN has a size of $32 \times 32 \times 32$. The Adam optimizer is used that allows to adjust the learning rate during the training phase. The learning rate $\lrate$ is set to $0.001$ to avoid the loss of local minima. The DQL algorithm prefers the long-term reward, and thus the discount factor $\dfactor$ is set to $0.9$. We use the $\epsilon$-greedy policy with $\epsilon=0.9$ that balances between the exploration and exploitation. During the training phase, $\epsilon$ is linearly reduced to zero that moves from the exploration to the exploitation. The probabilities of energy consumption and mobility of the workers are set as follows $p_l=p_{en}$ and $q_l=q_{mo}, \forall l \in \clL$. Other parameters are shown in Table~\ref{table:parameters}.

\begin{table}[h!]
	\centering
	\caption{Parameter Settings}
	\begin{tabular}{p{4.5cm}||p{2.0cm}}
		\hline
		$L=N=e^{max}$ & $3$ 	\\     
		\hline
		$\Delta$ & 5 \\
		\hline
		$\{\mu_1,\mu_2,\mu_3, \mu_{out}\}$ & $\{0.1,0.2,0.3,0.8\}$ \\
		\hline
		$\{\lambda_2, \lambda_3 \}$ & $\{2, 3\}$ \\
		\hline
		$\{p_{su}, p_{su}^2, p_{su}^3 \}$ & $\{0.5, 0.95, 0.98\}$ \\
		\hline
		$\{\alpha_\clI,\alpha_c,\alpha_e\} $ & $\{3,1,1\}$ \\ 
		\hline
		$\{p_{en}, q_{mo}\}$ & $\{0.5, 0.8\}$ \\ 
		\hline
	\end{tabular}
	\label{table:parameters}
\end{table}

\begin{figure}[t!]
	\centerline{\includegraphics[width=0.45\textwidth]{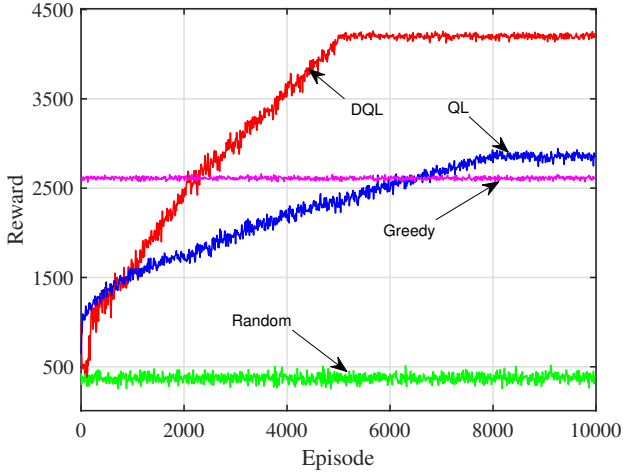}}
	\caption{Reward and convergence speed comparison.}
	\label{fig:DQL_compare}
\end{figure}

In Fig. \ref{fig:DQL_compare}, we plot the comparison among algorithms in terms of convergence speed and reward. As seen, the convergence speed of the DQL algorithm is much faster than that of the QL algorithm. Specifically, the DQL algorithm converges to the stable value of the reward within $5000$ episodes, while the QL algorithm needs to take around $8000$ episodes for the convergence. Also, the reward obtained by the DQL is much higher than those obtained by the baseline algorithms. In particular, the rewards obtained by the DQL, QL, greedy, and random algorithms are $4300$, $2700$, $2550$, and $480$, respectively. These results show that the DQL algorithm enables the model owner to learn the optimal polity. In particular with the greedy algorithm, the model owner employs the special channel with the highest quality and decides the maximum amount of energy recharged to each worker. The greedy algorithm can enable the model owner to improve its utility. However, it incurs the high channel cost and energy cost that significantly reduces the reward. For the random algorithm, the model owner randomly selects channels and amounts of energy. This may reduce the number of successful transmissions, and the workers may often face low energy states. Therefore, the random algorithm obtains the worst performance. 

\begin{figure}[t!]
	\centerline{\includegraphics[width=0.45\textwidth]{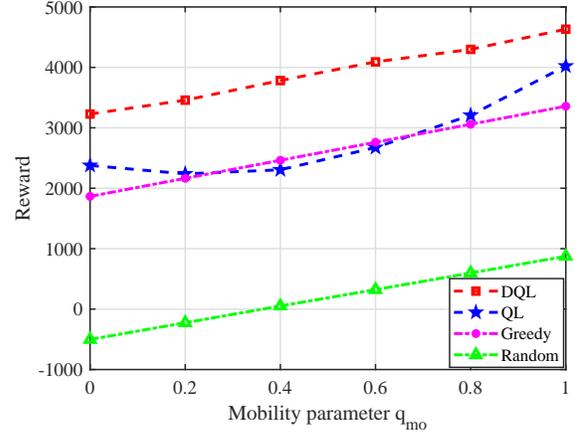}}
	\caption{The reward versus mobility parameter.}
	\label{fig:reward_qmo}
\end{figure}

The performance improvement of the DQL algorithm compared with the baseline algorithms is maintained even if the mobility parameter $q_{mo}$ varies as shown in Fig. \ref{fig:reward_qmo}. Note that as the mobility parameter increases, the rewards obtained by all the algorithms generally increase. This is due to the fact as $q_{mo}$ approaches $1$, most workers are in the coverage and the default channel is used. This significantly reduces the channel and energy costs and increases the reward. 


\begin{figure}[t!]
	\centerline{\includegraphics[width=0.45\textwidth]{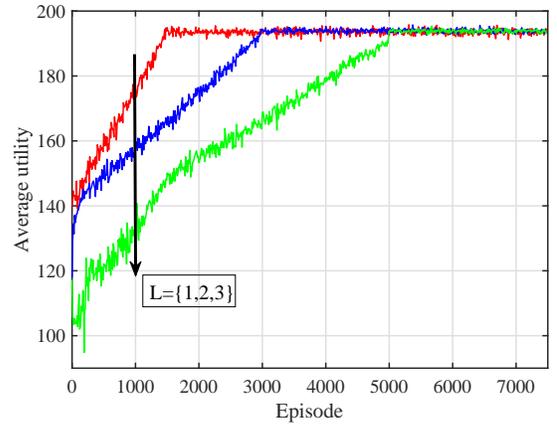}}
	\caption{The average utility with various $L$ values.}
	\label{fig:NoUser_incentive}
\end{figure}

Next, we evaluate the DQL algorithm as the number of workers $L$ varies.~Fig. \ref{fig:NoUser_incentive} shows the average utility obtained by the DQL algorithm under the different number of workers. As seen, the convergence speed of the DQL algorithm is slower as the number of workers increases. The reason is that as the number of workers increases, the action and state spaces increase that reduces the convergence speed of the algorithm. It is worth noting that the DQL algorithm converges to the same average utility regardless the number of workers. This is because of that the model owner already learns the optimal policy to obtain the maximum utility. 

\begin{figure}[t!]
	\centerline{\includegraphics[width=0.45\textwidth]{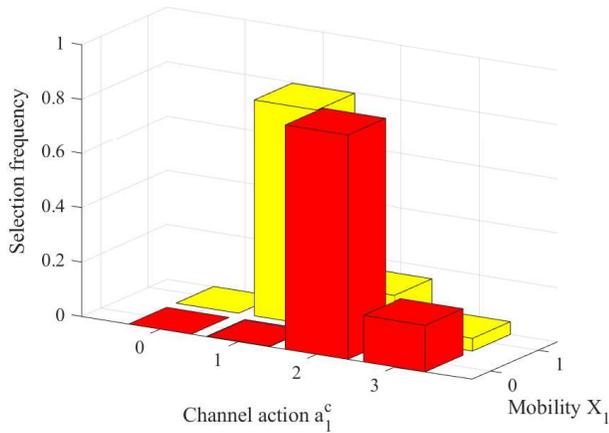}}
	\caption{Channel policy for worker 1.}
	\label{fig:3D}
\end{figure}

Finally, it is interesting to consider how the model owner selects the channel for transmitting its global model to each worker given the worker's mobility state. Without loss of
generality, we consider the worker 1. As shown in Fig. \ref{fig:3D}, as the worker is outside of the coverage, $X_1=0$, the model owner selects one special channel. Specifically, special channel $2$, i.e., $n=2$, is selected since it has a lower cost than other special channels. As the worker is present in the coverage, both the channels, i.e., the special channels and default channel, can be used. However, the selection frequency of the default channel is higher than that of the special channels. The reason is that the default channel has quality enough, e.g., $p_{su}=0.8$, and using the default channel results in reducing the channel cost.

\section{Conclusions}

In this paper, we have presented the DQL algorithm for the resource allocation in the mobility-aware federated learning network. In particular, we first formulate the channel selection and energy decision of the model owner as a stochastic optimization problem. The optimization problem aims to maximize the number of successful transmissions of the model owner while minimizing the energy and channel costs. To solve the problem, we have developed the DQL algorithm with DDQN. The simulation results show that the reward obtained by the proposed DQL is significantly higher than those obtained by the conventional algorithms. This means that the proposed DQL algorithm enables the model owner to learn the optimal decisions under the stochastic and uncertainty of the network environment.   

\bibliographystyle{IEEEtran}
\bibliography{IEEEabrv,Paper_Bib}

\begin{thebibliography}{10}
\providecommand{\url}[1]{#1}
\csname url@samestyle\endcsname
\providecommand{\newblock}{\relax}
\providecommand{\bibinfo}[2]{#2}
\providecommand{\BIBentrySTDinterwordspacing}{\spaceskip=0pt\relax}
\providecommand{\BIBentryALTinterwordstretchfactor}{4}
\providecommand{\BIBentryALTinterwordspacing}{\spaceskip=\fontdimen2\font plus
\BIBentryALTinterwordstretchfactor\fontdimen3\font minus
  \fontdimen4\font\relax}
\providecommand{\BIBforeignlanguage}[2]{{%
\expandafter\ifx\csname l@#1\endcsname\relax
\typeout{** WARNING: IEEEtran.bst: No hyphenation pattern has been}%
\typeout{** loaded for the language `#1'. Using the pattern for}%
\typeout{** the default language instead.}%
\else
\language=\csname l@#1\endcsname
\fi
#2}}
\providecommand{\BIBdecl}{\relax}
\BIBdecl

\bibitem{jing2017crowdtracker}
Y.~Jing, B.~Guo, Z.~Wang, V.~O. Li, J.~C. Lam, and Z.~Yu, ``Crowd{T}racker:
  {O}ptimized urban moving object tracking using mobile crowd sensing,''
  \emph{IEEE Internet Things J.}, vol.~5, no.~5, pp. 3452--3463, Oct. 2017.

\bibitem{mao2017survey}
Y.~Mao, C.~You, J.~Zhang, K.~Huang, and K.~B. Letaief, ``A survey on mobile
  edge computing: {T}he communication perspective,'' \emph{IEEE Commun. Surveys
  Tuts.}, vol.~19, no.~4, pp. 2322--2358, Aug. 2017.

\bibitem{mcmahan2016communication}
H.~B. McMahan, E.~Moore, D.~Ramage, S.~Hampson \emph{et~al.},
  ``Communication-efficient learning of deep networks from decentralized
  data,'' \emph{arXiv preprint arXiv:1602.05629}, Feb. 2016.

\bibitem{anh2019efficient}
T.~T. Anh, N.~C. Luong, D.~Niyato, D.~I. Kim, and L.-C. Wang, ``Efficient
  training management for mobile crowd-machine learning: {A} deep reinforcement
  learning approach,'' \emph{IEEE Wirel. Commun. Lett.}, May 2019.

\bibitem{mcmahan2017federated}
B.~McMahan and D.~Ramage, ``Federated learning: {C}ollaborative machine
  learning without centralized training data,'' \emph{Google Research Blog},
  vol.~3, Apr. 2017.

\bibitem{lim2019federated}
W.~Y.~B. Lim, N.~C. Luong, D.~T. Hoang, Y.~Jiao, Y.-C. Liang, Q.~Yang,
  D.~Niyato, and C.~Miao, ``Federated learning in mobile edge networks: {A}
  comprehensive survey,'' \emph{arXiv preprint arXiv:1909.11875}, Sep. 2019.

\bibitem{xu2016enterprise}
L.~Xu, J.~Xie, X.~Xu, and S.~Wang, ``Enterprise {LTE} and {W}i{F}i interworking
  system and a proposed network selection solution,'' in \emph{Proc. Symp.
  Archit. Netw. Commun. Syst.}, Mar. 2016, pp. 137--138.

\bibitem{afzal2018unlocking}
M.~K. Afzal, Y.~B. Zikria, S.~Mumtaz, A.~Rayes, A.~Al-Dulaimi, and M.~Guizani,
  ``Unlocking {5G} spectrum potential for intelligent {I}o{T}: {O}pportunities,
  challenges, and solutions,'' \emph{IEEE Commun. Mag.}, vol.~56, no.~10, pp.
  92--93, Oct. 2018.

\bibitem{mnih2015human}
V.~Mnih, K.~Kavukcuoglu, D.~Silver, A.~A. Rusu, J.~Veness, M.~G. Bellemare,
  A.~Graves, M.~Riedmiller, A.~K. Fidjeland, G.~Ostrovski \emph{et~al.},
  ``Human-level control through deep reinforcement learning,'' \emph{Nature},
  vol. 518, no. 7540, p. 529, Feb. 2015.

\bibitem{van2016deep}
H.~Van~Hasselt, A.~Guez, and D.~Silver, ``Deep reinforcement learning with
  double {Q}-learning,'' in \emph{AAAI}, Phoenix, Arizona, Mar. 2016, pp. 1--7.

\bibitem{tran2019federated}
N.~H. Tran, W.~Bao, A.~Zomaya, and C.~S. Hong, ``Federated learning over
  wireless networks: Optimization model design and analysis,'' in \emph{IEEE
  INFOCOM 2019-IEEE Conference on Computer Communications}, Apr. 2019, pp.
  1387--1395.

\bibitem{watkins1992q}
C.~J. Watkins and P.~Dayan, ``Q-learning,'' \emph{Machine learning}, vol.~8,
  no. 3-4, pp. 279--292, May 1992.

\bibitem{anh2018deep}
T.~T. Anh, N.~C. Luong, D.~Niyato, Y.-C. Liang, and D.~I. Kim, ``Deep
  reinforcement learning for time scheduling in {RF}-powered backscatter
  cognitive radio networks,'' \emph{arXiv preprint arXiv:1810.04520}, 2018.

\bibitem{abadi2016tensorflow}
M.~Abadi, P.~Barham, J.~Chen, Z.~Chen, A.~Davis, J.~Dean, M.~Devin,
  S.~Ghemawat, G.~Irving, M.~Isard \emph{et~al.}, ``Tensorflow: {A} system for
  large-scale machine learning,'' in \emph{Proc. 12th USENIX Symp. Operating
  Syst. Des. Implementation}, 2016, pp. 265--283.

\end{thebibliography}

\end{document}